\documentclass[twocolumn,amsmath,amssymb]{revtex4}
\usepackage{subfigure}
\usepackage{here}
\usepackage{float}
\usepackage{wrapfig}
\usepackage{graphicx}
\usepackage{dcolumn}
\usepackage{bm}
\usepackage{amsmath}
\usepackage{amsfonts}
\usepackage{amssymb}
\usepackage{color}
\usepackage{mathrsfs}
\usepackage{url}
\makeatletter
\pacs{}

\begin{document}
\title{Structure, design,  and mechanics of a pop-up origami with cuts}
\author{Taiju Yoneda$^{1}$, Yoshinobu Miyamoto$^{2}$ and Hirofumi Wada$^{1}$} 
\affiliation{$^{1}$ Department of Physical Sciences, Ritsumeikan University, Kusatsu, Shiga 525-8577, Japan \\
$^{2}$ Department of Architecture, Aichi Institute of Technology, Toyota, 470-0392, Japan}

\begin{abstract}
The rotational erection system (RES) is  an origami-based design method for generating a three-dimensional (3D) structure from a planar sheet. 
Its rotational and translational kinematics are fully encoded in the form of prescribed cuts and folds that imply negative degrees of freedom. 
Here, we characterize the mechanical properties of a threefold symmetric elastic RES by combining finite element analysis and a physical experiment. 
We demonstrate that plate bending creates a physical route connecting the two energetically separated configurations, that is, flat and standing states, allowing RES to morph into a 3D shape via a snap-through transition. 
We quantify the energy barrier for the bistability, and indicate that it is independent of the entire  structure's span, but depends solely on its aspect ratio. 
We also indicate that the standing RES has a proper structural rigidity and resilience, owing to its unique self-locking mechanism, suggesting its superior load-bearing performance in a range of applications. 
The present study clarifies the basic actuation mechanism of an origami-based deployable structure extended with chiral patterned cuts, providing  a way to use the optimally designed RES in a range of man-made systems.
\end{abstract}
\maketitle

{\it Introduction --}
An origami-inspired design enables the creation of a 3D structure from a planar sheet or membrane~\cite{Demaine-Review-2011, Tachi-JMecDes-2003, Lebee-Review-2015, Schenk-PNAS-2013, Silverberg-Science-2014, Silverberg-NatMat-2015, Lechenault-PRL-2015, Yasuda-PRL-2015, Faber-Science-2018}. 
Because a folded 3D configuration involves self-contact and overlap, while the surface area must be conserved owing to the sheet inextensibility, its lateral extent usually diminishes so that the structure grows vertically. 
This intrinsic property of origami underlies the prominent foldability of large membranes, which are utilized in a range of natural~\cite{Kobayashi-PRSLB-1998, Mahadevan-Science-2005, Py-PRL-2007, Harrington-NatComm-2011, Forbes-JEntomolSoc-1926, Saito-PNAS-2017} and man-made systems~\cite{Miura-Book-1985, Bruton-OpenSci-2016, Kuribayashi-MatSciEng-2006}. 

In contrast, kirigami is a design method that makes a paper sheet highly stretchable by periodically introducing free boundaries, that is, cuts and holes~\cite{Qi-PRB-2014, Isobe-SciRep-2016}. 
In a typical kirigami  design, a sheet is free of any overlap or self-contact, but is mechanically monostable; its extended state is stable only in the presence of external loading. 
Thus, it is interesting to create a multi-stable geometric design in which a single flat sheet morphs into a stable 3D structure vertically, without losing its lateral extent. 
A similar, but more straightforward example can be determined in certain types of pop-up cards and lift-the-flap books. 
Strategies for forming 3D microstructures for photonics and flexible electronics are also based on similar concepts, which are designed to be actuated by global buckling compression~\cite{Zhang-Review-2017}.
Clearly, an elaborate combination of folds and cuts, called, "ori-kirigami" is a promising method for developing a novel class of shape-shifting materials~\cite{Nojima-JSME-2006, Saito-JIMSS-2011, Castle-PRL-2014, Neville-SciRep-2016, Holms-Review-2019, Liu-SoftMat-2020}.

 \begin{figure}[b]
 \includegraphics[width=0.90\linewidth]{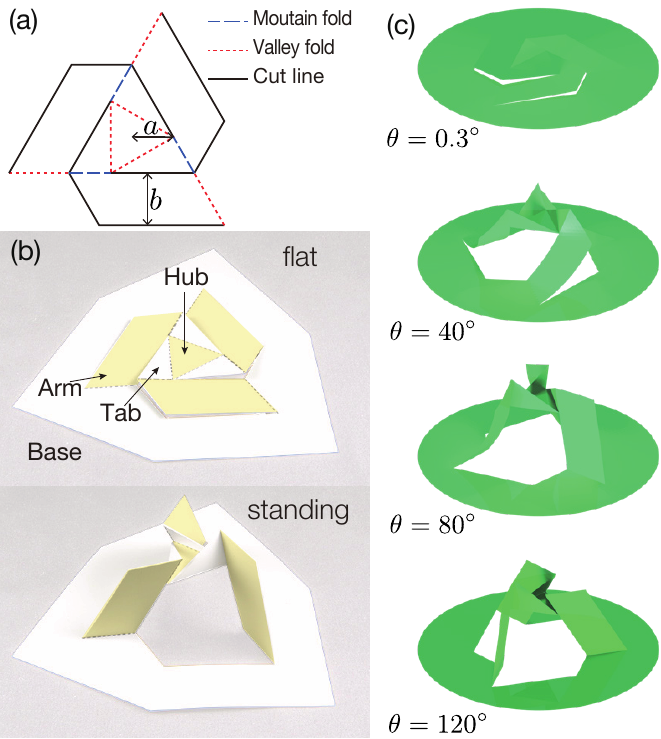}
 \caption{An overview of the rotational erection system (RES). (a) Mountain and valley creases, as well as cut lines, with the definition of geometric lengths $a$ and $b$.
 (b) Photographs of a paper-made RES in (almost) flat and standing configurations, with the definitions of the names of the specific parts. 
 (c) Sequence of snapshots of the RES generated from our finite element simulations (see the text for details)}.
 \label{fig:structure}
\end{figure}

The rotational erection system (RES) developed in Ref.~\cite{Miyamoto-6OSME-2014, Miyamoto-7OSME-2018, Miyamoto-flickr} provides a unique example of a 2D to 3D transformation without any global edge loading.
RES is bistable, and reversibly transforms into a stable 3D structure from a planar sheet with prescribed cuts and fold lines, only by local rotational actuation (Fig.~\ref{fig:structure}) .
The 3D standing shape is able to support an external load, as well as its own weight, implying that the RES would be an efficient production method for industrial, artistic, and architectural uses. 
However, a quantitative investigation of the structure and mechanics of the RES has been unexplored, limiting its practical applications. 
For example, snapping must be avoided for large-scale architectural purposes, whereas it may be exploited for switching devices, or RES-based energy-absorbing metamaterials. 
To control and optimize the RES functions, it is crucial to obtain a quantitative and predictive understanding of its  energy landscape and geometric scalability ,composed of real physical materials. 

To characterize the physical properties of the RES, we first conducted numerical simulations based on finite element analysis (FEA).
Compared to the existing origami simulators suitable for quickly grasping the folding behavior of an origami with a given crease pattern~\cite{Ghassaei-Sim-2018}, our FEA simulation is physically exact and fully quantitative because it accurately accounts for out-of-plane elastic bending and twisting, as well as in-plane stretching of all the plates constituting the entire RES.
Our FEA prediction is then quantitatively verified by  a physical experiment using a real RES engineered from a synthetic paper. 
 
 {\it Simulation and Experiments }
Although a variety of multi-stable RES designs exist, we focus on a threefold symmetric RES, as illustrated in Fig.~\ref{fig:structure} (a) that embodies 
an essential mechanism common to a family of more complex RES designs~\cite{comment-RES}. 
Because the RES morphs into a 3D shape as it is rotated in the vertical direction (defined as the $z$-axis), 
we characterize its configuration by its height, $z$, and rotational angle $\theta$~\cite{Miyamoto-6OSME-2014}.
Note that rotational actuation is a key element in the RES design, allowing it to take up additional lengths from a flat, inextensive sheet to grow vertically.
Throughout the experiment,  we applied an axial external torque to the RES to actuate it without any vertical external force.
For our physical RES comprising sufficiently stiff plates (explained below), the effects of gravity on the RES deformations are negligible. 

Structurally, RES comprises four parts, which include  the base, arms, tabs, and hub. (see Fig.~\ref{fig:structure}). 
A geometric RES comprising rigid surfaces joined by hinges has a negative degree of freedom (DOF), and is not rigidly foldable. 
Refer to  supplemental material (SM)~\cite{SM}. 

\begin{figure*}
 \includegraphics[width=0.97\linewidth]{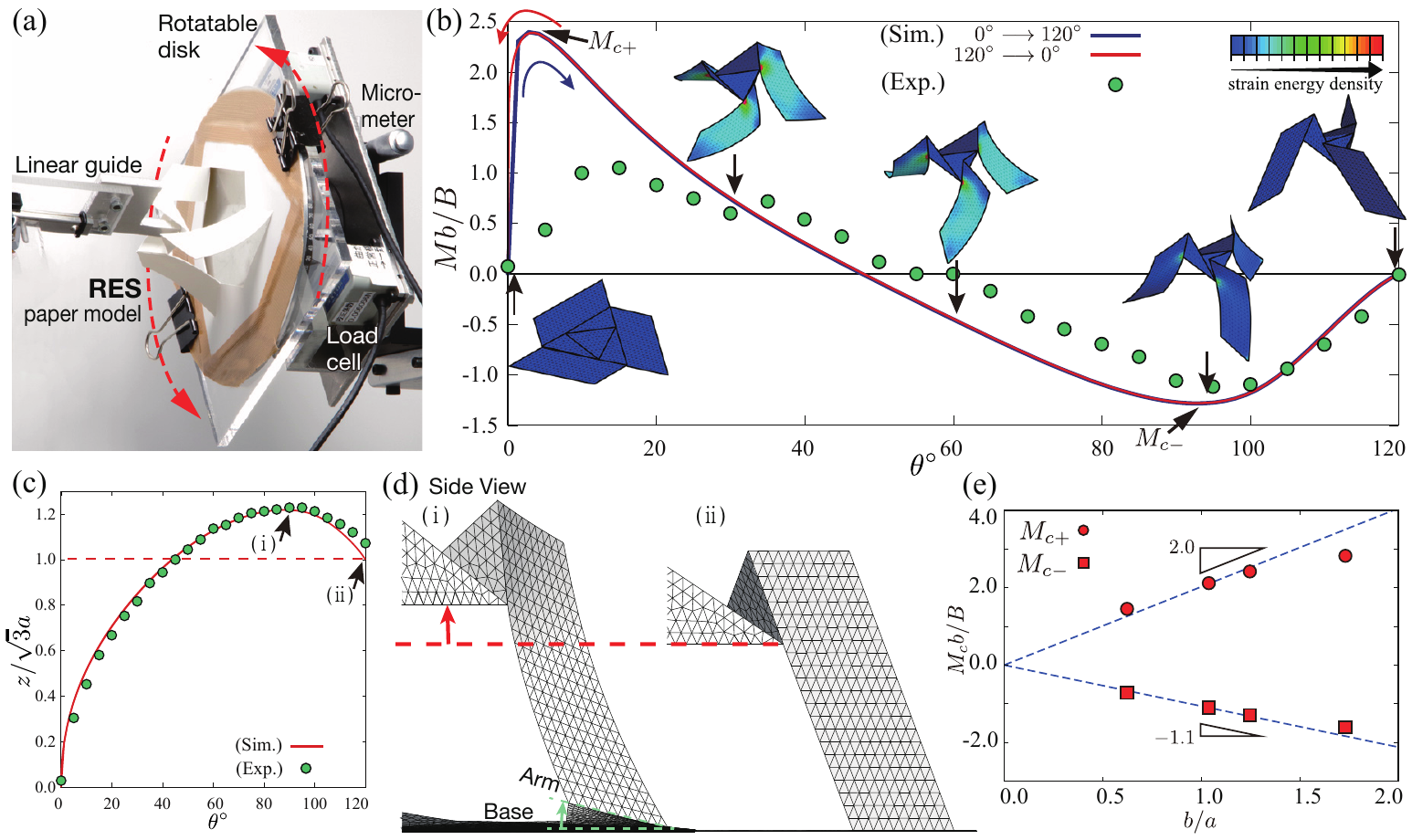}
 \caption{Experimental and numerical results for our RES mechanics and geometry. 
 (a) Experimental system for measuring the mechanical response of the RES.
 (b) Rescaled torque $Mb/B$ plotted as a function of rotational angle $\theta$ (in degrees) obtained from the experiment (filled green symbols), and the FEA simulations (solid red and blue lines). The corresponding FEA snapshots were displayed simultaneously . The color map illustrates the total strain energy density. 
 (c) Rescaled height of the RES (defined as the vertical distance between the base and the hub), $z/(\sqrt{3}a)$, plotted as a function of $\theta$, in a similar  way like (a). 
 (d) The snapshots in (i) and (ii) indicated in (b) were obtained from the FEA simulations. The horizontal red dotted line represents the equilibrium height of the hub in (ii). 
 (e) Peak values of the torque in (a) (indicated by $M_{c +}$ and $M_{c -}$) are a function of $a/b$. Dashed lines are fits to the data, based on the prediction of the scaling theory (refer to  the text).
 }
 \label{fig:torque}
\end{figure*}

Contrary to the prediction solely based on the geometry of the RES design, a shape transition of the RES made from real materials can be readily actuated by twisting the hub with respect to the base. 
To quantify this unique morphing pathway, we conducted an FEA simulation using the commercial software ABAQUS (Dassault Systems).
First, we generated the panels constituting the RES separately, which were then assembled with freely rotatable joints at ridges to build a complete RES. 
We modeled the base as a thin circular disc with radius $r$, which was attached to another larger disc that mimics the rotatable disc in our experiment, as explained below (Fig.~\ref{fig:torque} (a)).
For the actuation of an entire RES, this outermost disc is gradually rotated by a given step of angle quasi-statically.
The equilibrium configuration was obtained by minimizing the elastic energy of a linear isotropic solid using triangular shell elements with geometric nonlinearity.
Different values of the Young's modulus $E=20~\mathrm{MPa} - 2~\mathrm{GPa} $ and thickness $t=0.125--1.0\mathrm{mm}$ were examined, with a fixed Poisson's ratio $\nu = 0.3$.
The geometric parameters $a$ and $b$ were set to be equal to those in the experimental model given below. 
We systematically tested different sets of mesh sizes and types, as well as the base size $r$ (in the range of 60-80 mm), and confirmed that the presented results are essentially insensitive to those parameters.
We started our computation with a slightly lifted shape from a flat plane, and simulated a complete cycle. 

To verify our simulation results, we also conducted a physical experiment. 
Using our home-built system, we experimentally quantified the axial torque $M$ applied to the RES as a function of $\theta$. 
The RES with dimensions $ a = 25/\sqrt{3} ~ \mathrm{mm}$ and $b = 18 ~ \mathrm{mm}$ was processed at Fuji Toso Kogyo Co., Ltd. using a $t=0.5$ mm thick YUPO synthetic paper.
Because a YUPO paper has a well-defined anisotropy in its bending stiffness owing to the manufacturing process, 
we conducted an independent mechanical test for YUPO paper, to determine its direction-averaged bending stiffness per unit width of 0.022 N$\cdot$m~\cite{SM}.
This amounts to an effective Young's modulus of $E=1.9$ GPa, given the Poisson ratio $\nu=0.3$, which may be valid for the major component of YUPO, that is, polypropylene.)
We introduced a half-cut on a crease line that can be regarded as a freely rotatable hinge, compared to the typical forces required to deform the RES panels~\cite{comment-fab}.

For the torque measurement, we developed an experimental system that comprised  two load cells and a rotatable disc onto which an RES paper model was mounted (Fig.~\ref{fig:torque} (a)). 
The edge of the base was firmly attached to the disc with an adhesive tape, while  the hub plate was attached to a linear translational guide that prohibited any rotation.
Thus, the device allows the RES to take an arbitrary rotational angle $\theta$, without translational constraints.
The resulting tangential force $F$ necessary to maintain the entire structure at a given $\theta$ was measured using the load cells, from which we deduced  the torque by $M=FR$, where $R=60$ mm is the distance between the axis of rotation and the load cell. 
In the experiment, 
we investigated the folding (i.e., standing to flat) process only, in which the twisting angle $\theta$ was quasi-statically decreased by $5^{\circ}$, from $120^{\circ}$.
The mechanical response of the deployment process (flat to standing) was explored using FEA.

{\it Results--}
In Fig.~\ref{fig:torque}(b), we plot the rescaled experimental torque $Mb/ B$, where $B = Et^3 b/[12(1-\nu^2)]$ is the bending modulus per arm of width $b$~\cite{Audoly-Book}, together with the FEA simulation data. 
Overall, the numerical and experimental data exhibit a similar trend, confirming the validity of the two independent approaches. 

We first focus on the $\theta \ll 1$ region, where the RES is activated from the flat configuration.
For an ideal RES studied in the simulation, the s change in shape begins by buckling the three arms. 
Accordingly, the torque increases discontinuously at $\theta=0$, from zero to $M_{c+} \approx 2.5 B/b$ (Fig.~\ref{fig:torque} (b)]. 
Suppose a thin strip of length $L$, thickness $t$, and width $b$, subjected to an external force $f$ is applied at one end~\cite{Audoly-Book}. 
The strip buckles at $f_c\approx \alpha B/L^2$, where $\alpha$ is a numerical prefactor of the order of $1-10$ that depends sensitively on the boundary conditions and
geometry of the strip~\cite{comment-buckling}.
Assuming threefold symmetry, the buckling torque is $M_{c+} \sim 3f_ca \sim 3\alpha Ba/L^2$. 
For the arm of RES considered here, $L\approx 2a$, We predict $M_{c+}b/B \approx (3\alpha/4)b/a$.
In Fig.~\ref{fig:torque} (e), we plot $M_{c+}$ for different values of $b/a$ obtained from the simulations, which confirms that $M_{c+}b/B \sim b/a$. 
A deviation is solely observed for $b/a>1$, for which a narrow strip assumption is no longer valid. 
From fitting  the data in Fig.~\ref{fig:torque} (c), we determine that $\alpha \simeq 2.7$, which is quite a reasonable value as the prefactor of the critical buckling force~\cite{comment-buckling}. 
In contrast to the simulated RES, the arms of our physical RES are not entirely flat, but have some small permanent curvatures once the RES completes a cycle of shape transformation. 
Therefore, the physical RES can stand up from the (nearly) flat configuration without any buckling, leading to a smooth torque curve, as well as a reduced peak value, as illustrated  in Fig.~\ref{fig:torque} (b)].
At the maximum torque, the slope becomes negative with increasing $\theta$, and at $\theta=\theta_c \approx 60^{\circ}$, the torque crosses the unstable equilibrium point at which $M=0$.
The RES would jump to another stable configuration at $\theta=\theta_c$. 

We now focus on its behavior around the standing configuration at $\theta=120^{\circ}$.
To highlight the extent of departure from the standing state, we define $\varphi$ as $\theta=120^{\circ}-\varphi$.
We observe in Fig.~\ref{fig:torque} (b)that the magnitude of the torque increases approximately linearly with $\varphi$, for $0<\varphi<30-40^{\circ}$.
Interestingly, the height of the RES also increases  for this range of $\varphi$ [Fig.~\ref{fig:torque} (c)]. 
As the hub begins rotation, its arms are bent and simultaneously pushed outwards.
Because the arms are tilted with respect to the flat base, the hub must be initially lifted upwards under such combined deformations. Refer to Fig.~\ref{fig:torque} (d) (i) and (ii).
Note that the base slightly leaves the ground during the process. 
The compliance of the base significantly lowers the risk of the RES damage, particularly in the vicinity of the joints to the base, but is not essential for the kinematics illustrated  in Fig.~\ref{fig:torque} (c)].
As will be discussed later, the unique kinematics that "locks" its 3D configuration underlies a structural rigidity of the RES investigated below.

{\it Discussion--}
In Fig.~\ref{fig:torque} (b), the torque response of the RES was indicated to be fully reversible. 
However, the energy landscape of the RES around $\theta=0$ is actually more complicated than expected from Fig.~\ref{fig:torque} (b).
Multiple kinematic pathways are allowed for RES to take on at $\theta=0^{\circ}$, most of which lead to an incomplete transformation. 
 For RES to get on the "right" path at $\theta=0$, all the three arms need to buckle towards the same direction synchronously, so that the hub starts moving vertically (either upwards or downwards), keeping its horizontality. 
If a vertical tensile force is applied to the hub, the up-down symmetry is broken externally, and the synchronized buckling of the three arms is readily induced, which is exactly the way we involuntarily play with the RES manually. 
Similarly, if the arms have a small initial curvature in a similar direction, this also assists the entire structure in deforming smoothly  without elastic instabilities, which is what was observed in the torque measurement experiment. 
Altogether, a weak distortion intrinsic to structures made of real physical materials, combined with visually guided manipulations manually, can drastically simplify the morphing energy landscape, which then enables the RES to perform as intended, with a stable and reproducible cyclic actuation.
This feature will be a fairly generic consequence that is probably valid for a wide class of functionally foldable thin structures, beyond the specific example studied here.

\begin{figure}
 \includegraphics[width=0.99\linewidth]{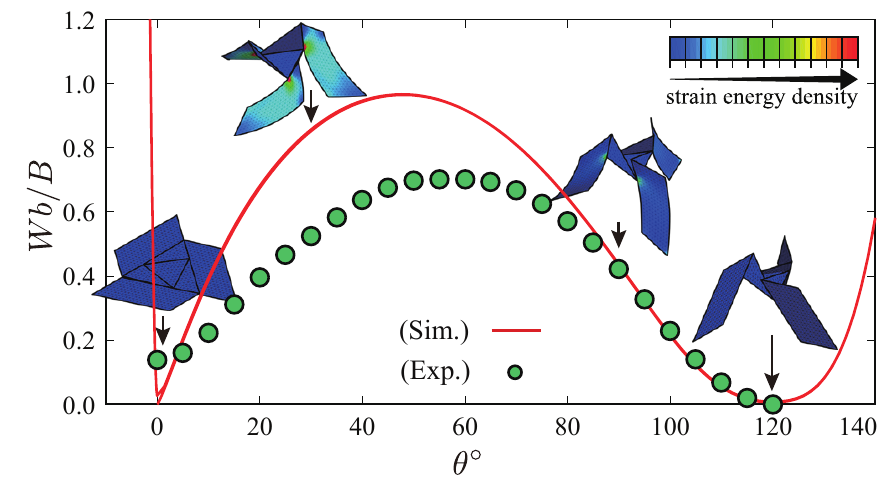}
\caption{The energy profiles for the bistability of RES, reconstructed from the torque vs. angle curves is illustrated in Fig.~\ref{fig:torque} (b), obtained from the simulation and experiment.
In the simulation data, the parameter range of $\theta$ was extended to include $\theta<0^{\circ}$ and $\theta>120^{\circ}$.}
\label{fig:energy}
\end{figure}

It is now evident that the RES jumps from one stable state to another by bending its three arms elastically. 
A distributed curvature over the surface of the arm works as a "virtual crease," much as for origami with a square-twist pattern~\cite{Silverberg-NatMat-2015}.
The bending energy connects a kinematic pathway between the flat and standing configurations that are otherwise isolated in the configuration space~\cite{Liu-NatPhys-2018}. 
We can quantify the energy barrier for bistability based on the torque response illustrated in Fig.~\ref{fig:torque} (b), according to $W=\int M(\theta)d\theta$, which is plotted in Fig.~\ref{fig:energy} as a function of $\theta$.
We reproduce an expected double-well  potential with two stable states of a similar energy value. 
According to the scaling argument given previously, we conclude that the energy barrier for the shape transition, $U$, is defined as 
\begin{eqnarray}
 U &\sim& \frac{Et^3b}{a}.
 \label{eq:U}
\end{eqnarray}
Remarkably, $U$ is independent of the entire span of the structure, but solely depends on the ratio $b/a$, indicating that the bistability of RES is essentially scale-independent. 
This confirms the validity of the bar-hinge (truss) model, in which a smooth deflection of the arm is substituted with a localized bending stiffness across the "virtual crease" independent of the length of the arm $a$.

\begin{figure}
\includegraphics[width=0.99\linewidth]{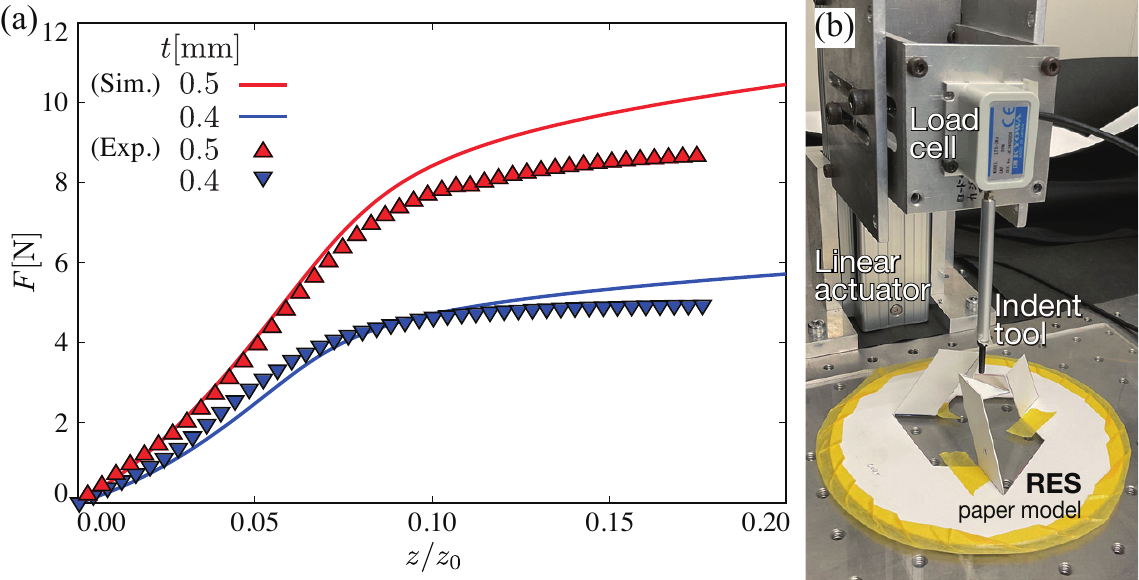}
\caption{Compression test of the RES in the standing configuration. (a) Load vs. displacement curves obtained from the experiment (symbols), and the FEA simulations (solid lines) for different plate thicknesses $t=0.4$ mm (blue) and 0.5 ~ mm (red). (b) Experimental setup for the force measurement. }
\label{fig:RES_force}
\end{figure}

\begin{figure}[b]
\includegraphics[width=0.78\linewidth]{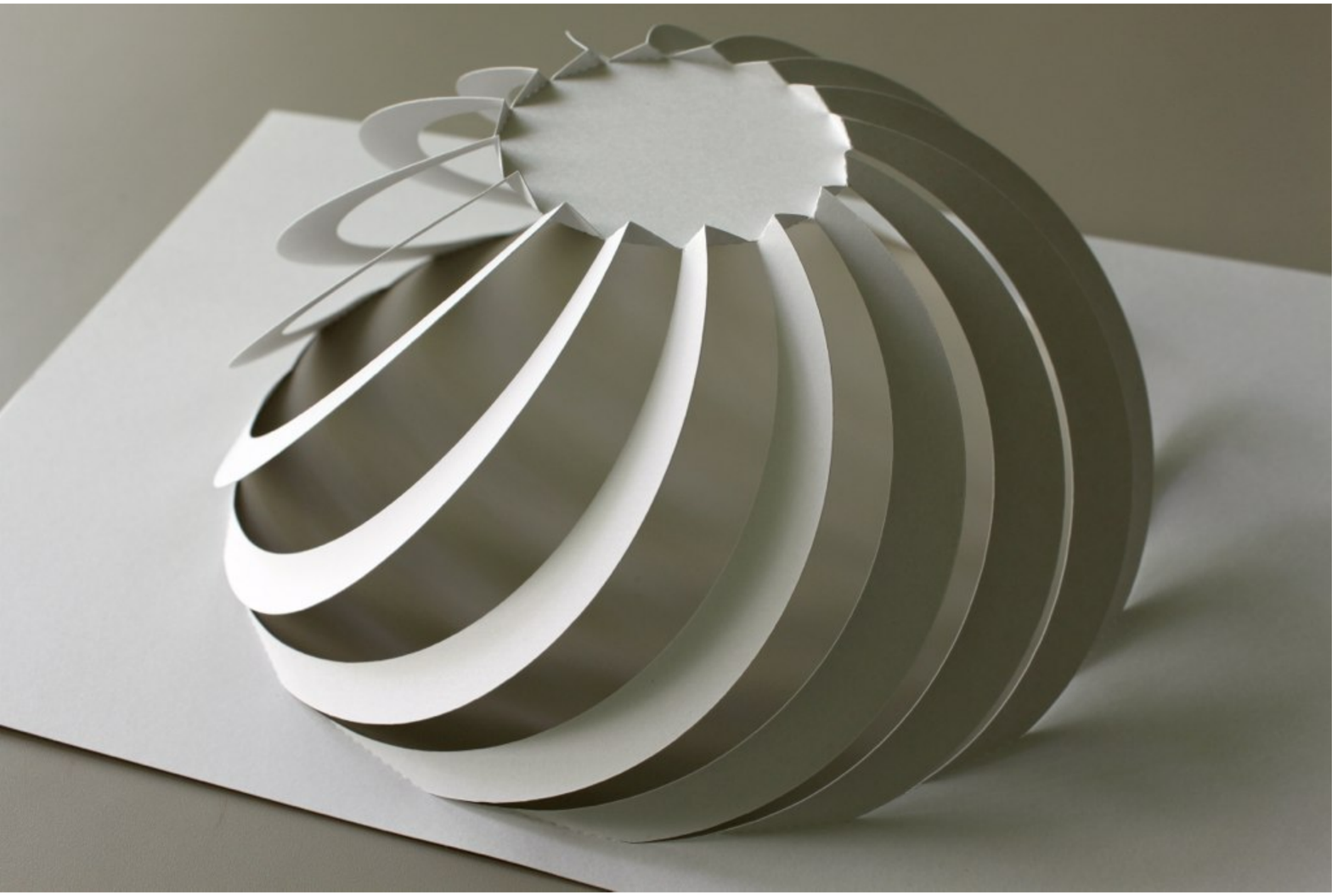}
\caption{A paper model of an  RES-based hemi-spherical dome. A collection of pleasing pictures of a variety of RES-based designs can be determined in Ref.~\cite{Miyamoto-flickr}.}
\label{fig:RESdome}
\end{figure}

To investigate the mechanical stability of our RES, we conducted a compression test to obtain the load-displacement curves (Fig.~\ref{fig:RES_force}).
We determined that a vertical compression force induces the twisting of the RES beyond $\theta=120^{\circ}$, thereby rigidifying the entire structure. 
Thus, the RES resists the flattening transition, recovering its original configuration immediately  the loading is removed, like an ordinary elastic spring ~\cite{Chang-CHI-2020}.
The restoring force increases with an increase in the vertical displacement, where the arms buckle and deflect correspondingly. 
Actually, the flattening transition (i.e., collapse) of the RES never occurs with compressional loading only. 
This is understood based on the kinematics of the RES illustrated  in Fig.~\ref{fig:torque} (c), where the RES first has to increase its height to finally reach the flat configuration. 
It has to be pulled upward first (with its rotation to be allowed freely), and then pushed downward, clearly an impossible mode with a pure compressive load only.
The built-in self-locking mechanism is illustrated in Fig.~\ref{fig:torque} (c), which underlies the structural rigidity of the standing RES in Fig.~\ref{fig:RES_force} (a)].

The actuation  of a rotationally patterned 3D structure from a planar sheet represents a robust design principle in a range of engineering systems~\cite{Zhang-Review-2017}. 
A variety of elaborate planar patterns of cuts and folds generate a family of RES with different 3D architectures and topologies, including domes (Fig.~\ref{fig:RESdome}) and multistage tower-like configurations as well ~\cite{Miyamoto-6OSME-2014}. For more, refer to ~\cite{Miyamoto-flickr}.
These complex designs share similar geometric and mechanical principles that has been revealed for the simplest threefold symmetric model here.
The scale-free and tunable nature of RES bistability will be potentially useful in a range of applications, including switching devices, energy-adsorbing mechanical systems, and one-step construction in architectures. 
Our proof-of-concept study clarifies the basic actuation mechanism of an origami-based deployable structure extended with cuts, and will determine the use of optimally designed RES in a range of artificial systems. 

\begin{acknowledgments}
Financial support from JSPS KAKENHI (Grant No. 18K13519 (to H.W.) No. 20K12518 (to Y. M.)) and Grant-in-Aid for JSPS Research Fellow (DC1, No. 19J22381 (to T.Y.)) is acknowledged.
\end{acknowledgments}


\begin{thebibliography}{38}
\bibitem[{com({\natexlab{a}})}]{comment-RES}
\bibinfo{note}{To be exact, our threefold symmetric RES is tri-stable. It can
  pop out either upwards or downwards, where resulting structures are
  mirror-symmetric, with their chiralities being opposite, and all the vally
  and mountain folds being interchanged. As those reflection-symmetric
  strucutres are mechanically identical, we describe in this study our RES as
  being bistable.}
\bibitem[{com({\natexlab{b}})}]{comment-fab}
\bibinfo{note}{Owning to small (but unavoidable) imperfections in the
  fabrication, the rotational axis of the hub and of the base did not match
  completely, and some strains remained in the both stable configurations,
  which is unimportant to our main results.}
\bibitem[{com({\natexlab{c}})}]{comment-buckling}
\bibinfo{note}{For a planar buckling of a uniform rod, we have $\alpha=\pi^2,
  4\pi^2$ and $\pi^2/4$, for hinged, clamped, and free boundary conditions,
  respectively. For more general boundary conditions and specific strip
  geometry including our case, the exact value of $\alpha$ will also fall
  within the range of $\pi^2/4-4\pi^2$.}
\end{thebibliography}

\end{document}